\documentclass[aps,prb,twocolumn,superscriptaddress]{revtex4}%
\usepackage{graphics}
\usepackage{amsmath}
\usepackage{graphicx}
\usepackage{amsfonts}
\usepackage{amssymb}
\usepackage{float}
\usepackage{longtable}
\usepackage{epsfig}
\usepackage{latexsym}
\usepackage{theorem}
\usepackage{bbm}
\usepackage{bm}
\usepackage{psfrag}
\begin{document}
\title{Advances in Photonic Quantum Sensing}
\author{Stefano Pirandola}
\affiliation{Computer Science and York Centre for Quantum
Technologies, University of York, York YO10 5GH, UK}
\affiliation{Research Laboratory of Electronics, MIT, Cambridge MA
02139, USA}
\author{Bhaskar Roy Bardhan}
\affiliation{Department of Physics and Astronomy, State University of New York at Geneseo,
Geneseo NY 14454, USA}
\author{Tobias Gehring}
\affiliation{Department of Physics, Technical University of Denmark, Fysikvej, 2800
Kongens, Lyngby, Denmark}
\author{Christian Weedbrook}
\affiliation{Xanadu, 372 Richmond St W, Toronto, M5V 2L7, Canada}
\author{Seth Lloyd}
\affiliation{Research Laboratory of Electronics, MIT, Cambridge MA 02139, USA}
\affiliation{Department of Mechanical Engineering, MIT, Cambridge MA 02139, USA}

\begin{abstract}
Quantum sensing has become a mature and broad field. It is generally related
with the idea of using quantum resources to boost the performance of a number
of practical tasks, including the radar-like detection of faint objects, the
readout of information from optical memories or fragile physical systems, and
the optical resolution of extremely close point-like sources. Here we first
focus on the basic tools behind quantum sensing, discussing the most recent
and general formulations for the problems of quantum parameter estimation and
hypothesis testing. With this basic background in our hands, we then review
emerging applications of quantum sensing in the photonic regime both from a
theoretical and experimental point of view. Besides the state-of-the-art, we
also discuss open problems and potential next steps.

\end{abstract}
\maketitle

Quantum technologies are today developing at unprecedented pace. As a matter
of fact, the technological applications of the field of quantum
information\cite{Nielsen,Hayashi,hybrid,serale,WeeRMP} are many and promising.
One of the most advanced areas is certainly quantum sensing. This is a broad
term which encompasses all those quantum protocols of estimation and
discrimination of physical parameters which are able to exceed the performance
of any classical strategy. Quantum sensing improves a number of tasks,
including gravitational wave detection, astronomical observations,
microscopes, target detection, data readout, atomic clocks, biological probing
and so on.

As with most technologies, quantum physics offers improvements that in some
cases far exceed anything that can be done classically. The quantum version of
sensing is no exception. By exploiting fundamental laws of physics one can
leverage important quantum characteristics such as entanglement, single
photons and squeezed states\cite{WeeRMP,SamPeter,Ger2} in order to achieve
orders-of-magnitude improvements in precision. In this scenario, the photonic
regime is certainly the best setting thanks to the relative simplicity in the
generation, manipulation and detection of such exotic quantum features.

This review aims to provide a survey of recent advances in photonic quantum
sensing. We refer the reader to Degen et al.\cite{Degen} for an overview of
quantum sensing in non-photonic areas, related to spin qubits, trapped ions,
and flux qubits. Here we start by providing mathematical background in quantum
parameter estimation\cite{Sam1,Sam2,Lore,Giova,ReviewNEW,SethSCIENCE,Paris}
and quantum hypothesis
testing\cite{Helstrom,QHT,QHT2,UNA1,QCB1,QCB2,QCB3,QHB1,Gae1,Invernizzi},
presenting the most general formulation of these problems. In fact, we
describe the most powerful (adaptive) protocols allowed by quantum mechanics
and how they can be reduced to simpler schemes by employing methods of channel
simulation~\cite{reviewMET,adaptivePRL}. This is an approach based on tools of
quantum programmability\cite{Prog1,Prog2,Prog3,Prog4} and teleportation
stretching\cite{PLOB,TQC}. This general background will allow us to identify
the goals, the structure, and the classical benchmarks for the following
protocols of quantum sensing that we will discuss both theoretically and experimentally.

Quantum hypothesis testing is at the very basis of quantum
reading\cite{Qread,Qread3,Qread4,Nair11,Hirota11,QreadCAP,Bisio11,Tej13,Arno12,Saikat2,ArnoIJQI,LupoSUPER,SaikatJeff,GaeENTROPY,Yen11}%
, where the information stored in an optical memory (or an equivalent physical
system) is retrieved by using extremely low energetic quantum states of light.
It is also at the basis of quantum
illumination\cite{Qillexp1,Qillexp2,Qillexp3,Meda,Qill0,Qill3,Qill1,Qill2,QillAMP,QillASY1,QillASY2,SaikatREC,Qillmetro,Ragy,JeffRECEIVER,ROC,LasHeras,JeffNJP}%
, where the radar-like detection of remote and very faint targets is boosted
by the use of quantum correlations. Quantum parameter estimation is the core
idea for the most recent advances in quantum imaging and optical
resolution\cite{Tsang15,Lupo16,Tsang2,Tang2016,Yang2016,Kerviche,Rehacek,Ran2016,Tham2017,Paur2016,Classen2016,Gatto,Treps,Yang17,Lu2018}%
, where \textquotedblleft Rayleigh's curse\textquotedblright\ may be dispelled
by using quantum metrological detection schemes\cite{Tsang15,Lupo16,Tsang2}
whose accuracy and sensitivity do not depend on the separation of two
point-like sources.

\smallskip

\textbf{Estimation and discrimination protocols}

\noindent Quantum sensing takes its roots in two problems which are central in
quantum information theory, known as quantum parameter estimation (or quantum
metrology) and quantum hypothesis testing (or quantum discrimination). A
modern formulation of these problems is respectively in terms of quantum
channel estimation and quantum channel discrimination, where the task is to
estimate or discriminate the values of a classical parameter encoded in a
quantum channel, i.e., a transformation between quantum states.
Mathematically, both the problems can be described by using an input-output
formalism where two parties, say Alice and Bob, probe a black box containing
the unknown quantum channel.

Therefore, consider a parameter $\theta$ encoded in a quantum channel
$\mathcal{E}_{\theta}$ and stored in a black box, of which Alice may prepare
the input and Bob may detect the output. In the estimation problem, $\theta$
is a continuous parameter, while in the discrimination problem, $\theta$ may
only take a discrete and finite number of values with some prior
probabilities. For the latter case, we consider here the most basic scenario:
binary symmetric discrimination, where $\theta$ may only take two values,
$\theta_{0}$ (null hypothesis) or $\theta_{1}$\ (alternative hypothesis), with
the same Bayesian cost and prior probability. This is then equivalent to
retrieving the classical bit $u$\ which is encoded in the parameter
$\theta_{u}$.

Let us analyze the estimation/discrimination problem with an increasing level
of complexity. In a basic \textquotedblleft block unassisted\textquotedblright%
\ protocol, Alice prepares an input state $\rho$ which is transmitted through
the unknown channel $\mathcal{E}_{\theta}$ and transformed into the unknown
output state $\mathcal{E}_{\theta}(\rho)$ for Bob. Consider this process to be
performed $n$ times, so that Alice sends $n$ copies $\rho^{\otimes n}$ and Bob
receives $\mathcal{E}_{\theta}(\rho)^{\otimes n}$ assuming that the channel is
memoryless. To retrieve $\theta$, Bob applies an optimal generally-joint
measurement to his $n$-copy output state. The type of measurement is different
depending on the problem. In channel estimation, the measurement has a
continuous outcome from which Bob constructs an unbiased estimator
$\tilde{\theta}$\ of $\theta$, affected by some error variance $\delta
\theta^{2}:=\langle(\tilde{\theta}-\theta)^{2}\rangle$. In channel
discrimination, Bob uses a dichotomic measurement which provides the bit $u$
with some mean error probability $p_{\text{err}}$.

\begin{figure*}[th]
\vspace{-2.2cm}
\par
\begin{center}
\vspace{-2.0cm} \includegraphics[width=0.99\textwidth]{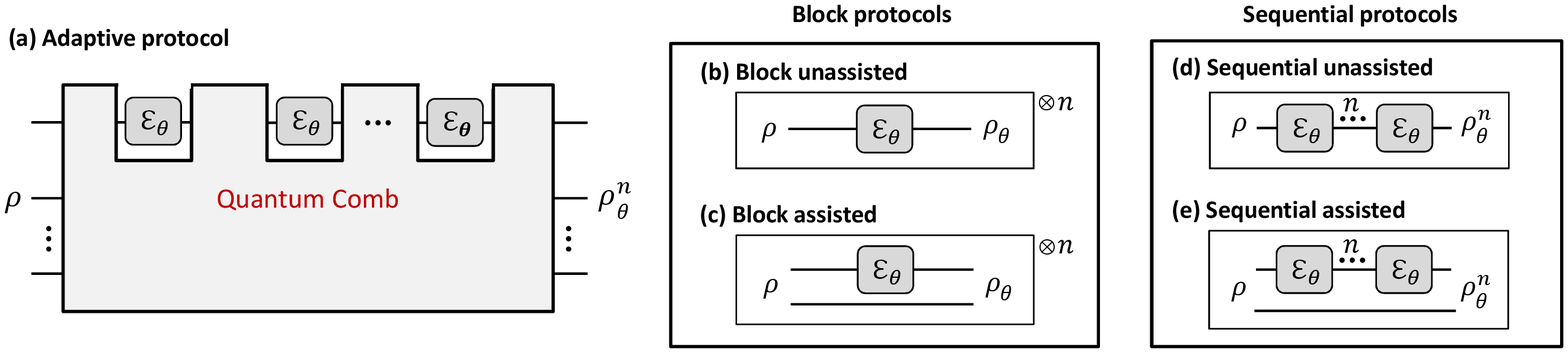}
\vspace{-4.5cm}
\end{center}
\caption{Protocols for quantum estimation and discrimination. \textbf{(a)~}%
Adaptive protocol represented as a quantum comb. An input register with an
arbitrary number of systems (wires) is prepared in a fundamental initial state
$\rho$. Each probing of the unknown channel $\mathcal{E}_{\theta}$\ is
performed by inputting a system from the register and storing the output back
in the register. Probings are interleaved by arbitrary QOs performed over the
entire register. After $n$ probings, the total output $\rho_{\theta}^{n}$\ is
subject to a joint quantum measurement. \textbf{(b)} Block unassisted protocol
where channel $\mathcal{E}_{\theta}$ is probed $n$ times in an identical and
independent way. \textbf{(c)}~Block assisted protocol where channel
$\mathcal{E}_{\theta}$ is probed by a signal system coupled to a reference
system.~\textbf{(d)}~Sequential unassisted protocol where the input is
transmitted through $n$ consecutive instances of the channel $\mathcal{E}%
_{\theta}$. \textbf{(e)}~Sequential assisted protocol where the input is
bipartite and partially transmitted through $n$ consecutive instances of
$\mathcal{E}_{\theta}$.}%
\label{qcombFIG}%
\end{figure*}

The most general estimation/discrimination protocol is based on unlimited
entanglement and adaptive quantum operations (QOs), which are applied jointly
by Alice and Bob\cite{adaptivePRL,Lore,Prog4,Harrow}. As also discussed in
ref.~\onlinecite{reviewMET}, this protocol can be represented as a quantum
comb~\cite{qcomb}. This is a quantum circuit board whose slots are filled with
the unknown quantum channel $\mathcal{E}_{\theta}$. The comb is based on a
register with an arbitrary number of systems, prepared in some fundamental
state $\rho$. The entire register undergoes arbitrary QOs before and after
channel $\mathcal{E}_{\theta}$ is probed, as depicted in Fig.~\ref{qcombFIG}.
The QOs can always be assumed to be trace-preserving by adding extra systems
and adopting the principle of deferred measurement~\cite{Nielsen}. At the
output of the comb, the state $\rho_{\theta}^{n}$ is detected by an optimal
quantum measurement whose outcome is classically processed.

It is clear that the quantum comb encompasses the previous block protocol with
output $\rho_{\theta}^{n}=\mathcal{E}_{\theta}(\rho)^{\otimes n}$. It also
includes the \textquotedblleft block-assisted\textquotedblright\ protocol
where the input state is bipartite and comprises a signal system $s$, sent to
probe the channel, and an idler or reference system $r$, which only assists
the output measurement. By repeating the process $n$ times, the output is
given by $\rho_{\theta}^{n}=\mathcal{E}_{\theta}\otimes\mathcal{I}(\rho
_{sr})^{\otimes n}$, where $\mathcal{I}$ is an identity channel. Finally, the
comb can also describe \textquotedblleft sequential\textquotedblright%
\ protocols where the input state is transmitted through $n$ instances of the
channel $\mathcal{E}_{\theta}\circ\cdots\circ\mathcal{E}_{\theta}$. See
Fig.~\ref{qcombFIG} for this classification.

\smallskip

\textbf{Performance of channel estimation}

\noindent Assume that the quantum comb in Fig.~\ref{qcombFIG} is used and
optimized for the problem of quantum channel estimation. Then, the ultimate
performance is limited by the quantum Cramer Rao bound (QCRB)%
\begin{equation}
\delta\theta^{2}\geq\frac{1}{\mathrm{QFI}(\rho_{\theta}^{n})}, \label{formSAM}%
\end{equation}
where $\mathrm{QFI}$ is the quantum Fisher information~\cite{Sam1}%
\begin{equation}
\mathrm{QFI}(\rho_{\theta}^{n})=\frac{8[1-F(\rho_{\theta}^{n},\rho
_{\theta+d\theta}^{n})]}{d\theta^{2}}, \label{QFIandFID}%
\end{equation}
with $F(\rho,\sigma):=\mathrm{Tr}\sqrt{\sqrt{\sigma}\rho\sqrt{\sigma}}$ being
the Bures fidelity between two states $\rho$ and $\sigma$.

We are interested in the \textquotedblleft scaling\textquotedblright\ of the
QCRB, i.e., on how $\delta\theta^{2}$ behaves for large number of uses $n$.
Here there are two main behaviors to consider. The first one is known as the
standard quantum limit (SQL) which is the typical scaling $\delta\theta
^{2}\gtrsim n^{-1}$ achievable in classical strategies. The other is the
Heisenberg limit \ $\delta\theta^{2}\gtrsim n^{-2}$ which can only be achieved
in a purely quantum setting. These have energy analogues when we consider
continuous variable systems\cite{WeeRMP} and parameter estimation with bosonic
channels. Assuming a single-use ($n=1$) of the comb and an energy-constrained
input state with $N$ mean number of photons, then $\delta\theta^{2}\gtrsim
N^{-1}$ corresponds to the SQL, while $\delta\theta^{2}\gtrsim N^{-2}$ is the
Heisenberg limit.

For instance, the Heisenberg limit is achievable for large $n$ when we
estimate the phase factor $\varphi$ of a unitary $U_{\varphi}=e^{i\varphi H}$
with $H$ being the free Hamiltonian. In fact, it is sufficient to use the
sequential protocol with an input pure state $|0,1\rangle_{sr}+|1,0\rangle
_{sr}$ so that we have the output $|0,1\rangle_{sr}+e^{in\varphi}%
|1,0\rangle_{sr}$ where the phase coherently accumulates. The same limit is
achievable in the number of photons $N$ by using the N00N state $|N,0\rangle
_{sr}+|0,N\rangle_{sr}$ in a single use of the block-assisted protocol.

We know a simple criterion to establish if channel estimation is limited to
the SQL. In particular, Pirandola and Lupo\cite{adaptivePRL} shed light on the
role of teleportation\cite{teleBennett,Samtele,telereview} in quantum
metrology finding that the property of teleportation covariance\cite{PLOB}
implies the SQL. Recall that a channel $\mathcal{E}$ is tele-covariant if, for
any teleportation unitary $U$ (Pauli\cite{Nielsen} or displacement
operator\cite{WeeRMP}), we may write\cite{PLOB}
\begin{equation}
\mathcal{E}(U\rho U^{\dagger})=V\mathcal{E}(\rho)V^{\dagger}, \label{teleCOV}%
\end{equation}
for a generally different unitary $V$. Because of this property, a quantum
channel is teleportation-simulable or Choi-stretchable, i.e., we may write the
simulation\cite{PLOB,TQC}%
\begin{equation}
\mathcal{E}(\rho)=\mathcal{T}(\rho\otimes\rho_{\mathcal{E}}), \label{telesim}%
\end{equation}
where $\mathcal{T}$ is the QO of teleportation and $\rho_{\mathcal{E}}$ is the
channel's Choi matrix $\rho_{\mathcal{E}}:=\mathcal{E}\otimes\mathcal{I}%
(\Phi_{sr})$, with $\Phi_{sr}$ being a maximally entangled state.

Here two observations are in order. First, let us remark that the simulation
in Eq.~(\ref{telesim}) needs a suitable asymptotic formulation\cite{PLOB,TQC}
for bosonic channels, whose Choi matrices are energy-unbounded. In fact, a
bosonic channel $\mathcal{E}$ has Choi matrix $\rho_{\mathcal{E}}:=\lim_{\mu
}\rho_{\mathcal{E}}^{\mu}$, where the Choi sequence $\rho_{\mathcal{E}}^{\mu
}:=\mathcal{E}\otimes\mathcal{I}(\Phi_{sr}^{\mu})$ is based on the two-mode
squeezed vacuum (TMSV) state\cite{WeeRMP} $\Phi_{sr}^{\mu}$ with variance
$\mu=2N+1$, with $N$ being the mean number of thermal photons in each mode.
Second, let us note that a teleportation-covariant channel $\mathcal{E}$ is a
specific type of programmable channel\cite{Prog1,Prog2,Prog3,Prog4}, where the
quantum gate array is teleportation and the program state is the Choi matrix
of the channel. The teleportation simulation has specific advantages in terms
of achievability of the QCRB\cite{reviewMET}.

By definition, a parametrized quantum channel $\mathcal{E}_{\theta}$ is
jointly teleportation-covariant\cite{adaptivePRL} if we may write
Eq.~(\ref{teleCOV}) for any $\theta$, i.e., $\mathcal{E}_{\theta}(U\rho
U^{\dagger})=V\mathcal{E}_{\theta}(\rho)V^{\dagger}$, so that the output
unitary $V$ depends on $U$ but not on $\theta$. Because of this property, we
may write the teleportation simulation $\mathcal{E}_{\theta}(\rho
)=\mathcal{T}(\rho\otimes\rho_{\mathcal{E}_{\theta}})$\ for any $\theta$.
Replacing this simulation in each slot of the quantum comb in
Fig.~\ref{qcombFIG} and \textquotedblleft stretching\textquotedblright\ the
adaptive protocol, we write the output state as%
\begin{equation}
\rho_{\theta}^{n}=\Lambda(\rho_{\mathcal{E}_{\theta}}^{\otimes n})~,
\label{simSTAND}%
\end{equation}
for a global quantum channel $\Lambda$. Because the QFI is monotonic under
channels and multiplicative over tensor products, one finds that
$\mathrm{QFI}(\rho_{\theta}^{n})\leq n\mathrm{QFI}(\rho_{\mathcal{E}_{\theta}%
})$, so that the QCRB must satisfy the SQL
\begin{equation}
\delta\theta^{2}\geq\lbrack n\mathrm{QFI}(\rho_{\mathcal{E}_{\theta}})]^{-1},
\label{reduction}%
\end{equation}
where we implicitly mean $\mathrm{QFI}(\rho_{\mathcal{E}_{\theta}}):=\lim
_{\mu}\mathrm{QFI}(\rho_{\mathcal{E}_{\theta}}^{\mu})$ for a bosonic channel.
In other words, the adaptive protocol has been reduced to a block assisted
protocol, where $n$ maximally-entangled states $\Phi_{sr}^{\otimes n}$ are
used to probe the unknown quantum channel $\mathcal{E}_{\theta}$.

Because the class of teleportation-covariant channels is wide, we have that
channel estimation is limited to the SQL in many situations. For instance, the
estimation of the probability parameter $p$ in depolarizing, dephasing or
erasure channels is limited to\cite{adaptivePRL} $\delta p^{2}\geq
p(1-p)n^{-1}$. Then, the estimation of the thermal noise $\bar{n}$ in a
thermal-loss channel $\mathcal{E}_{\eta,\bar{n}}$ with fixed transmissivity
$\eta$ is limited to\cite{adaptivePRL} $\delta\bar{n}^{2}\geq\bar{n}(\bar
{n}+1)n^{-1}$, which sets the limit for estimating excess noise in quantum key
distribution. By contrast, the ultimate estimation limit for the
transmissivity $\eta$ is not known. Because $\mathcal{E}_{\eta,\bar{n}}$ is
not jointly teleportation-covariant in $\eta$, the reduction in
Eq.~(\ref{reduction}) does not apply.

The optimal estimation of bosonic loss is an open problem with a number of
partial results. Solving this problem is important because loss is the main
decoherence effect in quantum optical communications, from fibre-based to
free-space implementations. Recall that the transmissivity $\eta$ limits the
ultimate rate achievable by point-to-point protocols of quantum and private
communication according to the Pirandola-Laurenza-Ottaviani-Banchi
bound\cite{PLOB}, so that $-\log_{2}(1-\eta)$ bits per use cannot be exceeded
without repeaters. Estimating $\eta$ is therefore of paramount importance. The
best-known performance in estimating the transmissivity $\eta$ of a pure-loss
channel $\mathcal{E}_{\eta}:=\mathcal{E}_{\eta,0}$ is limited to the SQL
$\delta\eta^{2}\geq\gamma N^{-1}$, with a pre-factor $\gamma$\ which has been
improved over time.

Monras and Paris\cite{Monras} studied the block unassisted protocol with
single-mode Gaussian states. They established the limit $\delta\eta^{2}%
\geq\eta N^{-1}$ achievable by using coherent-states, also known as shot-noise
limit or classical benchmark. They also found that squeezing may improve the
pre-factor and, therefore, beat the shot-noise limit. These results may be
achieved by computing the QFI as in Eq.~(\ref{QFIandFID}) and using the
formulas for the fidelity between Gaussian states\cite{Banchi}. The best
performance so far has been achieved by using the block protocol with
non-Gaussian states, including Fock states. In this way, Adesso et
al.\cite{AdessoOPT} found the improved scaling $\delta\eta^{2}\geq\eta
(1-\eta)N^{-1}$.


\smallskip

\textbf{Performance of channel discrimination}

\noindent Now assume that the quantum comb in Fig.~\ref{qcombFIG} is used and
optimized for binary discrimination, i.e., for the retrieval of the parameter
$\theta$ from a binary alphabet $\{\theta_{0},\theta_{1}\}$ where the two
possible values have identical Bayesian cost and the same prior probability.
This is now a problem of quantum channel discrimination, where we aim at
distinguishing between two channels $\mathcal{E}_{0}=\mathcal{E}_{\theta_{0}}$
and $\mathcal{E}_{1}=\mathcal{E}_{\theta_{1}}$, or equivalently at retrieving
the classical bit $u$ from $\mathcal{E}_{u}$. Let us call $\rho_{0}^{n}$ and
$\rho_{1}^{n}$ the two possible output states of the comb. The optimal
performance in terms of the minimum error probability is given by the Helstrom
bound\cite{Helstrom}
\begin{equation}
p_{\text{err}}(\mathcal{E}_{0},\mathcal{E}_{1})=[1-D(\rho_{0}^{n},\rho_{1}%
^{n})]/2,
\end{equation}
where $D(\rho,\sigma):=||\rho-\sigma||/2$ is the trace-distance\cite{Nielsen}.
This is obtained by a suitable dichotomic positive-operator valued measure
(POVM), called Helstrom POVM. Equivalently, note that the maximum classical
information $J$\ retrieved from the box is equal to%
\begin{equation}
J=1-H_{2}[p_{\text{err}}(\mathcal{E}_{0},\mathcal{E}_{1})],
\end{equation}
where $H_{2}$ is the binary Shannon entropy.

The error probability greatly simplifies if the two channels $\mathcal{E}_{0}$
and $\mathcal{E}_{1}$ are jointly teleportation-covariant, i.e., we may write
$\mathcal{E}_{u}(U\rho U^{\dagger})=V\mathcal{E}_{u}(\rho)V^{\dagger}$ for any
$u$. This allows us to use the teleportation simulation $\mathcal{E}_{u}%
(\rho)=\mathcal{T}(\rho\otimes\rho_{\mathcal{E}_{u}})$ over the Choi matrix
$\rho_{\mathcal{E}_{u}}$. We may then stretch the comb and write its output as
$\rho_{u}^{n}=\Lambda(\rho_{\mathcal{E}_{u}}^{\otimes n})$ for a global
channel $\Lambda$. Because the trace distance is monotonic under $\Lambda$, we
have $p_{\text{err}}\geq\lbrack1-D(\rho_{\mathcal{E}_{0}}^{\otimes n}%
,\rho_{\mathcal{E}_{1}}^{\otimes n})]/2$. Now note that this is achievable by
an assisted protocol which exploits maximally-entangled states at the input,
so that\cite{adaptivePRL}
\begin{equation}
p_{\text{err}}(\mathcal{E}_{0},\mathcal{E}_{1})=[1-D(\rho_{\mathcal{E}_{0}%
}^{\otimes n},\rho_{\mathcal{E}_{1}}^{\otimes n})]/2, \label{errMIN}%
\end{equation}
where $D=\lim_{\mu}D(\rho_{\mathcal{E}_{0}}^{\mu\otimes n},\rho_{\mathcal{E}%
_{1}}^{\mu\otimes n})$ for bosonic channels. In finite dimension,
Eq.~(\ref{errMIN}) implies that the diamond distance between two jointly
teleportation-covariant channels is simply equal to $||\mathcal{E}%
_{0}-\mathcal{E}_{1}||_{\diamond}=||\rho_{\mathcal{E}_{0}}-\rho_{\mathcal{E}%
_{1}}||$.

Starting from Eq.~(\ref{errMIN}), we may write simple lower and upper bounds
using the Fuchs-van de Graaf relations\cite{Fuchs} and the quantum Chernoff
bound (QCB)\cite{QCB1}. In fact, recall that for any pair of multicopy states
$\rho_{0}^{\otimes n}$ and $\rho_{1}^{\otimes n}$, the minimum error
probability $p_{\text{err}}=[1-D(\rho_{0}^{\otimes n},\rho_{1}^{\otimes
n})]/2$ satisfies the fidelity-based lower bound and the QCB%
\begin{align}
p_{\text{err}}  &  \geq\frac{1-\sqrt{1-F(\rho_{0},\rho_{1})^{2n}}}{2}%
:=F_{-}^{(n)}(\rho_{0},\rho_{1}),\label{BB1}\\
p_{\text{err}}  &  \leq\frac{Q(\rho_{0},\rho_{1})^{n}}{2},~~Q:=\inf
_{s}\mathrm{Tr}(\rho_{0}^{s},\rho_{1}^{1-s}). \label{BB2}%
\end{align}
In particular, for multimode Gaussian states, $\rho_{0}$ and $\rho_{1}$, we
know closed formulas for computing the fidelity\cite{Banchi} and the
QCB\cite{QCB3}. These inequalities can be extended to the adaptive error
probability of Eq.~(\ref{errMIN}) valid for jointly teleportation-covariant
channels, so that we may write\cite{adaptivePRL}%
\begin{equation}
F_{-}^{(n)}(\rho_{\mathcal{E}_{0}},\rho_{\mathcal{E}_{1}})\leq p_{\text{err}%
}(\mathcal{E}_{0},\mathcal{E}_{1})\leq\frac{Q(\rho_{\mathcal{E}_{0}}%
,\rho_{\mathcal{E}_{1}})^{n}}{2}, \label{boundsADA}%
\end{equation}
with asymptotic functionals over bosonic Choi matrices.

The results in Eqs.~(\ref{errMIN}) and~(\ref{boundsADA}) apply to many cases,
including the adaptive discrimination of Pauli channels, erasure channels, and
the noise parameters in bosonic Gaussian channels, such as the thermal photons
$\bar{n}_{0}$ and $\bar{n}_{1}$ in thermal-loss channels $\mathcal{E}%
_{\eta,\bar{n}_{0}}$ and $\mathcal{E}_{\eta,\bar{n}_{1}}$\ with the same
transmissivity $\eta$. Unfortunately, the same reduction does not apply to the
discrimination of bosonic loss, e.g., $\eta_{0}$ and $\eta_{1}$ of a
thermal-loss channel with fixed noise $\bar{n}$, because $\mathcal{E}%
_{\eta_{0},\bar{n}}$ and $\mathcal{E}_{\eta_{1},\bar{n}}$ are not jointly
teleportation-covariant. As a result, establishing the optimal discrimination
of bosonic loss is still an open problem. What we know currently is that
block-assisted strategies based on entangled states may greatly outperform
block unassisted strategies, especially when we employ a small number of
signal photons in the presence of thermal noise. This is the observation which
is at the basis of the applications of quantum reading and quantum illumination.

\smallskip

\textbf{Quantum reading of classical data}

\begin{figure}[th]
\vspace{-0.5cm}
\par
\begin{center}
\includegraphics[width=0.5 \textwidth]{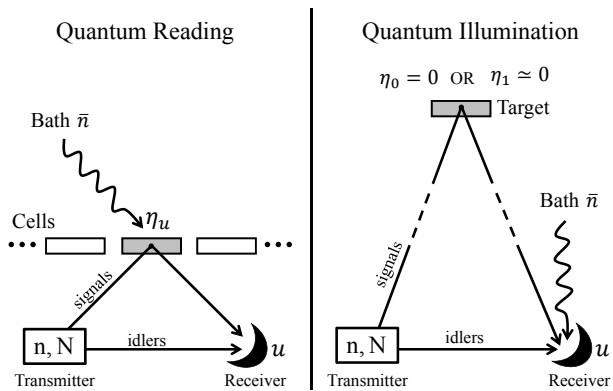}
\end{center}
\par
\vspace{-1cm} \caption{Quantum reading (left) and Gaussian quantum
illuminantion (right). In the basic formulation, these are both based on an
EPR transmitter, so that $n$ two-mode squeezed vacuum (TMSV) states irradiate
$N$ mean photons per mode over the cell/target (where $N$ is typically low).
The reflected signals are combined with the retained idler (reference) modes
in a joint detection, whose output $u$ discriminates between two hypotheses.
In quantum reading, $u$ is the information bit encoded into a cell with
reflectivity $\eta_{u}$ and subject to thermal noise $\bar{n}$. In quantum
illumination, $u$ is related with the absence ($\eta_{0}=0$) or the presence
($\eta_{1}\simeq0$) of a low-reflectivity target. The reflection is mixed with
an environment with bright thermal noise $\bar{n}\gg1$. These schemes are
examples of block-assisted protocols for quantum channel discrimination. In
the regimes considered, they largely outperform classical strategies, i.e.,
corresponding schemes based on classical transmitters that are not entangled
but composed of mixtures of coherent states.}%
\label{QreadFIG}%
\end{figure}

\noindent In 2011, ref.~\onlinecite{Qread} showed how the readout of classical
data from an optical digital memory can be modelled as a problem of quantum
channel discrimination. In the most basic description, an optical classical
memory can be seen as an array of cells described as microscopic beamsplitters
with different reflectivities (see Fig.~\ref{QreadFIG}). Each cell stores an
information bit $u=0,1$ in two equiprobable reflectivities, the
pit-reflectivity $\eta_{0}\in(0,1)$ and the land-reflectivity $\eta_{1}%
>\eta_{0}$. This single-cell model is equivalent to a black-box model read in
reflection so that the reflectivity plays the role of the transmissivity
parameter. The readout may also be affected by thermal noise, e.g., due to
stray photons generated by the source. Thus the readout corresponds to
discriminating between two thermal-loss channels, $\mathcal{E}_{0}%
:=\mathcal{E}_{\eta_{0},\bar{n}}$ and $\mathcal{E}_{1}:=\mathcal{E}_{\eta
_{1},\bar{n}}$, with different reflectivity, $\eta_{0}$ and $\eta_{1}$, but
fixed thermal number $\bar{n}$. Other decoherence effects may be
included\cite{Qread}, such as optical diffraction, memory effects and
inter-bit interference\cite{Qread4}.

We may consider different \textquotedblleft transmitters\textquotedblright%
\ composed of signal modes probing the cell and reference modes assisting
detection. The coherent-state transmitter\ only uses $n$ signal modes in
identical coherent states $\left\vert \alpha\right\rangle _{s}\left\langle
\alpha\right\vert ^{\otimes n}$. More powerfully, we may define a
\textquotedblleft classical\textquotedblright\ transmitter in the quantum
optical sense\cite{Prepres,Prepres2}. This is a block assisted protocol
employing mixtures of coherent states $\int d^{2n}\boldsymbol{\alpha
~}\mathcal{P}(\boldsymbol{\alpha})\left\vert \boldsymbol{\alpha}\right\rangle
\left\langle \boldsymbol{\alpha}\right\vert $, where $\mathcal{P}%
(\boldsymbol{\alpha})$\ is a probability distribution of amplitudes
$\boldsymbol{\alpha}$, and $\left\vert \boldsymbol{\alpha}\right\rangle
\left\langle \boldsymbol{\alpha}\right\vert $ is a multimode coherent state
with $n$ signal modes and $n$ reference modes. The optimal classical
transmitter has to be compared with an Einstein-Podolsky-Rosen (EPR)
transmitter. This is a block entanglement-assisted protocol where we send part
of $n$\ TMSV states $\Phi_{sr}^{\mu\otimes n}$, so that each signal mode is
entangled with a reference or \textquotedblleft idler\textquotedblright\ mode.

For both the classical and the EPR transmitter, we therefore assume an input
$2n$-mode state $\rho_{sr}$, which is transformed by the cell into an output
state $\sigma_{u}:=\mathcal{E}_{u}^{\otimes n}\otimes\mathcal{I}^{\otimes
n}(\rho_{sr})$ for the $n$ reflected signal modes and the $n$ kept reference
modes. This output is detected by an optimal Helstrom POVM with some error
probability. We then compare the optimal information retrieved by the
classical transmitter $J_{\text{class}}$ with that retrieved by the EPR
transmitter $J_{\text{EPR}}$, quantifying the information gain $\Delta
:=J_{\text{EPR}}-J_{\text{class}}$. Positive values\ $\Delta>0$ provide
quantum advantage. A fair comparison between these transmitters involves
fixing the mean number of signal photons probing the cell. Because each cell
is probed $n$ times, we may consider different types of energy constraints.

One type of constraint is \textquotedblleft
local\textquotedblright, meaning that we fix the mean number of
photons $N$ in each probing, so that the total energy scales as
$nN$. We may write the following general bound for the error
probability $p_{\text{class}}$ achievable by any classical
transmitter\cite{Qread}%
\begin{equation}
p_{\text{class}}\geq\mathcal{C}(n,N):=\frac{1-\sqrt{1-F(N)^{2n}}}{2},
\label{classBenchmark}%
\end{equation}
where $F(N)$ is Bures fidelity between the two possible outputs $\mathcal{E}%
_{0}(|\sqrt{N}\rangle\langle\sqrt{N}|)$ and $\mathcal{E}_{1}(|\sqrt{N}%
\rangle\langle\sqrt{N}|)$ generated by a single-mode coherent state with $N$
mean photons. This leads to the upper bound $J_{\text{class}}(n,N)\leq
1-H_{2}[\mathcal{C}(n,N)]$. For the EPR transmitter, we may exploit the QCB
which gives the lower bound $J_{\text{EPR}}\geq1-H_{2}[Q(\rho_{\mathcal{E}%
_{0}}^{\mu},\rho_{\mathcal{E}_{1}}^{\mu})^{n}/2]$, where $\rho_{\mathcal{E}%
_{u}}^{\mu}$ is the quasi-Choi matrix of the unknown channel $\mathcal{E}_{u}$
probed by a TMSV state $\Phi_{sr}^{\mu}$ with $\mu=2N+1$. Using these bounds
we construct a sufficient condition to prove the quantum advantage $\Delta>0$.

Numerical investigations show that positive values of $\Delta$ are typical for
low signal photons and high reflectivities in wide ranges for the thermal
noise. A positive quantum advantage can already be achieved by a single probe
per cell ($n=1$). When the land-reflectivity is very high $\eta_{1}%
\rightarrow1$ (ideal memory), one may derive analytical
expressions\cite{Qread3} and find regimes of parameters for which
$\Delta\rightarrow1$ bit per cell. This extremal value means that the
EPR\ transmitter is able to fully read the cell, while classical transmitters
do not retrieve any information. This advantage may also be used to design
cryptographic memories whose data can only be read by
entanglement\cite{GaeENTROPY}.

Another type of energy constraint is \textquotedblleft
global\textquotedblright, meaning that we fix the total mean number of photons
$N_{\text{T}}$ over all the $n$ uses, so that we employ an average of
$N_{\text{T}}/n$ photons per probing. Let us call $n$ the bandwidth of the
transmitter, due to the fact that this can be physically related with the
number of effective frequencies used in the readout. One can then show that,
at sufficiently low energies $N_{\text{T}}\lesssim10$ photons, a narrowband
EPR\ transmitter (even monochromatic $n_{\text{EPR}}=1$) is able to beat
arbitrary classical transmitters, up to extremely large bandwidths. Because a
few entangled photons can retrieve more information than any classical source
of light, we may indeed work at very low energies. This regime may be mapped
into much faster optical readers and denser memories\cite{Qread}.

Quantum reading has been extensively
studied\cite{Nair11,Hirota11,QreadCAP,Bisio11,Tej13,Arno12,Saikat2,ArnoIJQI,LupoSUPER,SaikatJeff,Yen11}%
. Ref.~\onlinecite{QreadCAP} extended the model to multi-cell error correction
coding and defined the notion of quantum reading capacity, a quantity that was
later shown to be super-additive\cite{LupoSUPER}. Guha and
Shapiro\cite{SaikatJeff} also defined a similar notion of
capacity\cite{Saikat2} and studied the error exponent for quantum reading. For
single-cell reading of an ideal memory in noiseless conditions,
Nair\cite{Nair11} showed that Fock states are optimal. More generally,
entangled states with the signal beam in a number-diagonal reduced state may
also provide a positive quantum advantage\cite{Nair11}. This class of states
is optimal for non-adaptive discrimination with single-mode and multi-mode
pure-loss channels\cite{Yen11}.

Hirota\cite{Hirota11} proposed an alternative model based on a binary phase
encoding and showed how entangled coherent states may achieve error-free
quantum reading. Non-Gaussian entangled states were also considered by Prabhu
Tej et al.\cite{Tej13}. Then, Bisio et al.\cite{Bisio11} studied a noise-free
unitary\ model of quantum reading where both the inputs of the unknown
beamsplitter are accessible for probing and both its outputs for detection.
Assuming a single probe ($n=1$), they found that the optimal two-mode input is
the superposition of a N00N state and the vacuum $\left\vert 00\right\rangle
$. This approach was later extended by Dall'Arno et al.\cite{Arno12} to
unambiguous quantum reading, where the probability of error is replaced by an
inconclusive result.

\begin{figure*}[th]
\vspace{-0.0cm}
\par
\begin{center}
\includegraphics[width=0.96 \textwidth]{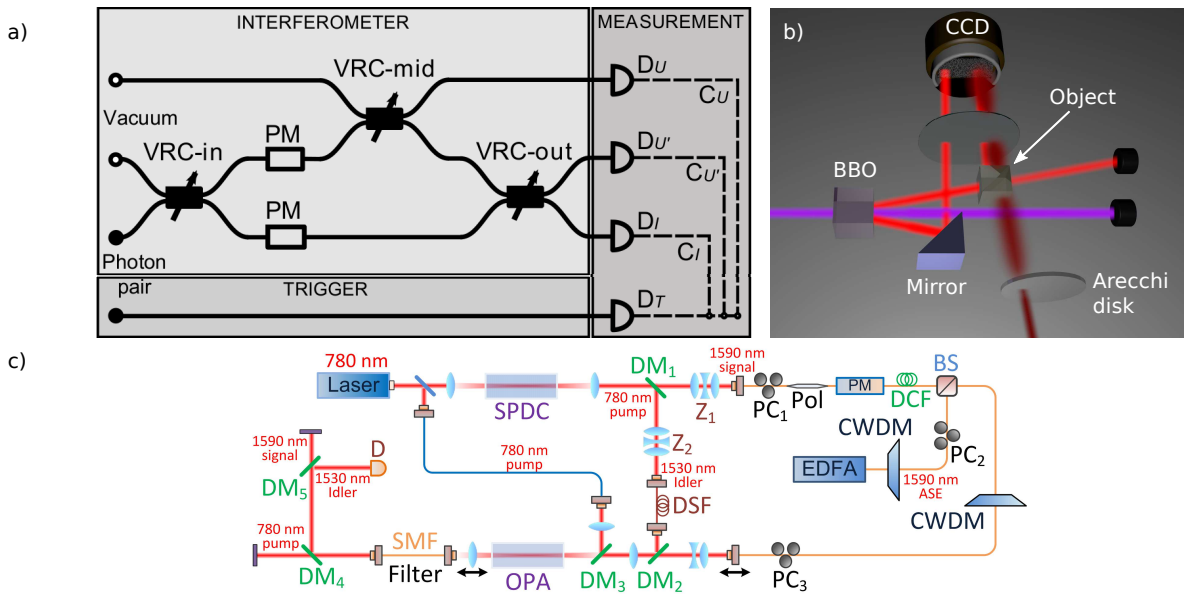}
\end{center}
\par
\vspace{-0.3cm}\caption{Experimental demonstrations of quantum
reading and quantum illumination. \textbf{a)} Experimental setup
of perfect quantum reading\cite{Arno12}. A photon pair source is
used to generate a heralded single photon using a trigger detector
($\mathrm{D}_T$). The heralded single photon is fed into an
interferometer with variable ratio couplers (VRCs) and phase
modulators (PMs) to add additional phase shifts. Coincidence
detection of the outputs are used to discriminate between two
possible splitting ratios of VRC-mid. \textbf{b)} Quantum
illumination experiment of Lopaeva et al.\cite{Qillexp1}. See also
Ref. \onlinecite{Meda}. Both beams of a photon pair source are
detected by a photon counting CCD camera. In the experiment the
target is a 50:50 beamsplitter placed in one of the beams. The
beamsplitter is simulated to be in a thermal environment by
illuminating it with scattered light from an Arrechi disk.
\textbf{c)} Quantum illumination experiment of Zhang et
al.\cite{Qillexp3}. Photon pairs generated by spontaneous
parametric down-conversion (SPDC) at two different wavelengths are
split using a dichroic mirror (DM). One of the photons is stored
in a delay line using a dispersion-shifted LEAF fiber (DSF). The
other photon is phase modulated (PM). A lossy and noisy
environment is simulated by a beamsplitter (BS) and amplified
spontaneous emission (ASE) from a erbium doped fiber amplifier
(EDFA). The joint detection is implemented using an optical
parameteric amplifier (OPA) whose output is detected by a PIN
photo detector (D). DCF: dispersion-compensating fiber, POL:
polarizer, CWDM: coarse wavelength-division multiplexer. Thin
lines are optical fiber, thick lines are unguided propagation.
Figures adapted with permission from: a),
ref.~\onlinecite{Arno12}, \copyright ~2012 APS; b),
ref.~\onlinecite{Qillexp1}, \copyright ~2013 APS; c)
ref.~\onlinecite{Qillexp3}, \copyright ~2015 APS.}%
\label{ExpREAD}%
\end{figure*}

Similar to Hirota\cite{Hirota11}, Dall'Arno et al.\cite{Arno12} also
considered a version of perfect quantum reading with zero discrimination
error. This is possible by designing an ideal cell which is either a
beamsplitter with perfect reflectivity ($\eta_{1}=1$) or a beamsplitter with
lower reflectivity $\eta_{0}<1$ and suitable $\pm\pi/2$\ phase shifters at the
input and output ports. This scheme was experimentally
implemented\cite{Arno12}. As shown in Fig.~\ref{ExpREAD}(a), the setup
consisted of a Mach-Zehnder interferometer with the variable beamsplitter
situated in one arm. A heralded single photon source based on spontaneous
parametric down-conversion in a $2$mm long beta barium borate (BBO) crystal
pumped with a continuous-wave laser diode at $405$nm served as the quantum
state source. The photon pairs generated in orthogonal polarizations in the
type-II phase matching process were separated by a polarizing beamsplitter.
While one photon was used to herald the process, the other was fed into the
Mach-Zehnder interferometer.

The setup discriminated between two beamsplitter configurations, the one with
reflectivity $\eta_{1}=1$ and the other with reflectivity $\eta_{0}<1$ and an
additional phase shift in the Mach-Zehnder interferometer arm. The measurement
consisted of three photon counters, one at the output of the variable
beamsplitter under test and the other two at each output of the
interferometer. Coincidence counts between each of the three detectors and the
trigger detector were measured using a $3$ns coincidence window. With the
perfect beamsplitter under test, only one of the detectors at the output of
the Mach-Zehnder interferometer would detect the photon (Hong-Ou-Mandel
effect), while for the beamsplitter with $\eta_{0}<1$ then any of the two
other detectors would detect the photon (due to the additional phase shift).

\smallskip

\textbf{Quantum illumination of targets}

\noindent Quantum sensing can be used not only to enhance the readout of
information from classical systems, but also to boost the standoff detection
of remote objects. This idea was first pushed forward by the efforts of Lloyd
and Shapiro at MIT\cite{Qill0,Qill1,JeffNJP}. In 2008, Lloyd\cite{Qill0}
designed a qubit-based protocol of quantum illumination, showing how the
detection of a low-reflectivity target object can be enhanced by using quantum
entanglement. The advantage of the entangled transmitter over non-entangled
ones is achieved even if the entanglement itself is completely lost after
reflection from the target. In fact, the initial signal-idler entanglement is
mapped into residual but yet quantum correlations between the reflected signal
and the kept idler that a suitably-designed quantum detector may
\textquotedblleft amplify\textquotedblright\ with respect to the uncorrelated
thermal background.

In the same year, Shapiro's team\cite{Qill1} proposed a practical formulation
of quantum illumination based on continuous-variable systems\cite{WeeRMP}.
Ref.~\onlinecite{Qill1}\ designed a Gaussian version where bosonic modes are
prepared in Gaussian states and sent to detect an object with low reflectivity
$\eta\simeq0$ in a region with bright thermal noise, i.e., with $\bar{n}\gg1$
mean thermal photons. The detection process can be modelled as the
discrimination between a zero-reflectivity thermal-loss channel $\mathcal{E}%
_{\eta=0,\bar{n}}$ (target absent) and a low-reflectivity thermal-loss channel
$\mathcal{E}_{\eta,\bar{n}^{\prime}}$\ with $\eta\simeq0$ and $\bar{n}%
^{\prime}=\bar{n}/(1-\eta)$ (target present). Here the factor $(1-\eta)^{-1}$
excludes the possibility of a \textquotedblleft passive
signature\textquotedblright\ which means the possibility of detecting the
target without transmitting any radiation by just measuring a lower received
background level. As also depicted in Fig.~\ref{QreadFIG}, one assumes that
the detector's noise does not depend on the presence of the target.

In this setup, we assume a local energy constraint, so that $N$ mean photons
are irradiated by each of the $n$ bosonic modes sent over the target object.
Under this assumption, we compute the error probability associated with the
various transmitters. In particular, we exploit the bounds in Eqs.~(\ref{BB1})
and~(\ref{BB2}) to compare the performance of the EPR transmitter (based on
TMSV states) with that of the classical transmitter (based on coherent
states). In the regime of low-energy signals ($N\ll1$) and many modes ($n\gg
1$), the EPR transmitter has the scaling\cite{Qill1} $p_{\text{EPR}%
}^{\text{err}}\simeq\exp(-n\eta N/\bar{n})/2$, which clearly beats the
classical transmitter $p_{\text{class}}^{\text{err}}\geq\exp(-n\eta N/2\bar
{n})/4$. In particular, $p_{\text{EPR}}^{\text{err}}$ realizes a 6dB advantage
in the error-probability exponent over the coherent-state transmitter
$p_{\text{CS}}^{\text{err}}\simeq\exp(-n\eta N/4\bar{n})/2$. Zhuang et
al.\cite{JeffRECEIVER} proved that the theoretical limit $p_{\text{EPR}%
}^{\text{err}}$ can be achieved by an explicit quantum receiver based on
feed-forward sum-frequency generation. This receiver has been also used to
show the quantum illumination advantage in terms of detection probability
versus false-alarm probability\cite{ROC}.

In 2015, Gaussian quantum illumination was extended to the microwave regime,
thus providing a prototype of quantum radar\cite{Qill2}. In this scheme, an
electro-optomechanical converter\cite{EOM1,EOM2,EOM3} transforms an optical
mode into microwave. If this transducer has high quantum efficiency, then
optical-optical entanglement is translated into microwave-optical
entanglement. The microwave signal is sent to probe the target region, while
the optical idler is retained. The microwave radiation collected from the
target region is then phase conjugated and upconverted into an optical field
by a second use of the transducer. The optical output is finally combined with
the retained idler in a joint detection, following the ideas behind the
receiver design by Guha and Erkmen\cite{SaikatREC}. In this way,
ref.~\onlinecite{Qill2} found that the error probability of microwave quantum
illumination is superior to that of any classical radar of equal transmitted
energy. A followup analysis has been carried out by Xiong et al.\cite{QillAMP}.

More recently, Sanz et al.\cite{Qillmetro} studied the protocol of quantum
illumination using the tools of quantum metrology so as to measure the
reflectivity of the target. They employed the QFI to bound the error
probability showing a 3dB-enhancement of the signal-to-noise ratio with
respect to the use of local measurements. They also considered non-Gaussian
Schr\"{o}dinger's cat states. Other studies have quantified the quantum
illumination advantage in terms of \textquotedblleft
consumption\textquotedblright\ of discord\cite{Discord} associated with the
target\cite{Qill3}, and in terms of mutual information\cite{Ragy}. Finally
note that quantum illumination has been also studied as an asymmetric Gaussian
discrimination problem by various authors\cite{Gae1,ROC,QillASY1,QillASY2}. In
this setting, TMSV states have been identified as optimal
probes\cite{QillASY2}. Finding the ultimate performance achievable by an
\textit{adaptive} version of quantum illumination remains an open question.

Several experimental implementations of quantum illumination have been
reported\cite{Qillexp1,Qillexp2,Qillexp3}. As depicted in Fig.~\ref{ExpREAD}%
(b), Lopaeva et al.\cite{Qillexp1} exploited a parametric down-conversion
source using a BBO crystal to generate two intensity-correlated light beams in
orthogonal polarizations at 710nm. Both beams were detected by a
photon-counting high-quantum efficiency CCD camera. The target object, a 50:50
beamsplitter in the experiment, was placed in one of the two entangled beams
before detection. The beamsplitter object was illuminated by photons scattered
on an Arecchi's rotating ground glass to simulate a thermal environment. A
single captured image was used to measure the second-order correlations
between the two beams. The implementation shows robustness against noise and
losses, and demonstrates a quantum enhancement in target detection in thermal
environments even when nonclassicality is lost. However, coincidence detection
of spontaneous parametric down-conversion is not the optimal detection method
to extract the most information from the signal-idler entangled modes, and the
implemented classical scheme using weakly thermal states is also non-optimal.

Adopting a different approach, in 2013 Zhang et al.\cite{Qillexp2} reported a
secure communication experiment based on quantum illumination. More recently,
Zhang et al.\cite{Qillexp3} demonstrated the advantage of quantum illumination
over coherent states by using broadband entangled Gaussian states, as produced
by continuous-wave spontaneous parametric down-conversion. In the experiment
shown in Fig.~\ref{ExpREAD}(c), the signal modes were phase modulated before
probing the weakly-reflecting target, while the idler modes were stored in a
delay line. The joint measurement was performed by combining the reflected
signal modes and the idler modes with a pump in another optical parametric
amplifier. The output in the order of nW was then detected by a PIN photo
detector with high gain and low noise. They showed a 20\% improvement of the
signal-to-noise ratio in comparison to the optimal classical scheme in an
environment exhibiting 14dB loss and a thermal background 75dB above the
returned probe light.

\smallskip

\textbf{Optical resolution beyond the Rayleigh limit}

\noindent The Rayleigh criterion is a well-known result in classical imaging.
Two point-like sources cannot be optically resolved (in the far field) if they
are closer than the Rayleigh length $\simeq\lambda/a$, where $\lambda$ is the
wavelength of the emitted light and $a$ is the numerical aperture of the
observing lens. For this reason, if we use a converging optical system to
focus light on a screen and an array of detectors to measure the intensity,
the Rayleigh's criterion together with the presence of photon shot noise, can
lead to severe limitations in resolving point-like sources. The situation
changes completely if we consider a fully quantum description of the light and
the measurement apparatus. In this way, Tsang et al.\cite{Tsang15} showed the
existence of a quantum detection scheme able to measure the distance between
two point-like sources with a constant accuracy, even when the sources have
sub-wavelength separation. This ground-breaking result was achieved by
relating the problem of resolving two incoherent point-like sources to quantum
estimation theory and using the QCRB.

The theory beyond these results were extended from incoherent sources emitting
faint pulses to thermal sources of arbitrary brightness\cite{Tsang2,Lupo16}.
In general, ref.~\onlinecite{Lupo16} established a connection between optical
resolution and bosonic channel estimation, so that measuring the separation
between two point-like sources is equivalent to estimating the loss parameters
of two lossy channels. In this way, ref.~\onlinecite{Lupo16} developed a
theory of super-resolution for point-like sources emitting light in a generic
state, i.e., attenuated or bright, classical, coherent, incoherent, as well as
entangled (e.g., in a microscope setup). The ultimate resolution was found as
a function of the optical properties of the two sources and their separation.
In particular, super-resolution can be enhanced when the sources emit
entangled or quantum-correlated (discordant) light.

More recently, Kerviche et al.\cite{Kerviche} extended Tsang et al.'s analysis
from a Gaussian point spread function to a hard-aperture pupil, proving the
information optimality of image-plane sinc-Bessel modes. They also generalized
the result to an arbitrary point spread function. Rehacek et al.\cite{Rehacek}
carried out further work on the optimal measurements for beating the Rayleigh
limit. Yang et al.\cite{Yang17} explored the use of homodyne or heterodyne
detection. Finally, Lu et al.\cite{Lu2018} studied the quantum-optimal
detection of one-versus-two incoherent optical sources with arbitrary separation.

\begin{figure*}[th]
\vspace{-0.0cm}
\par
\begin{center}
\includegraphics[width=0.95 \textwidth]{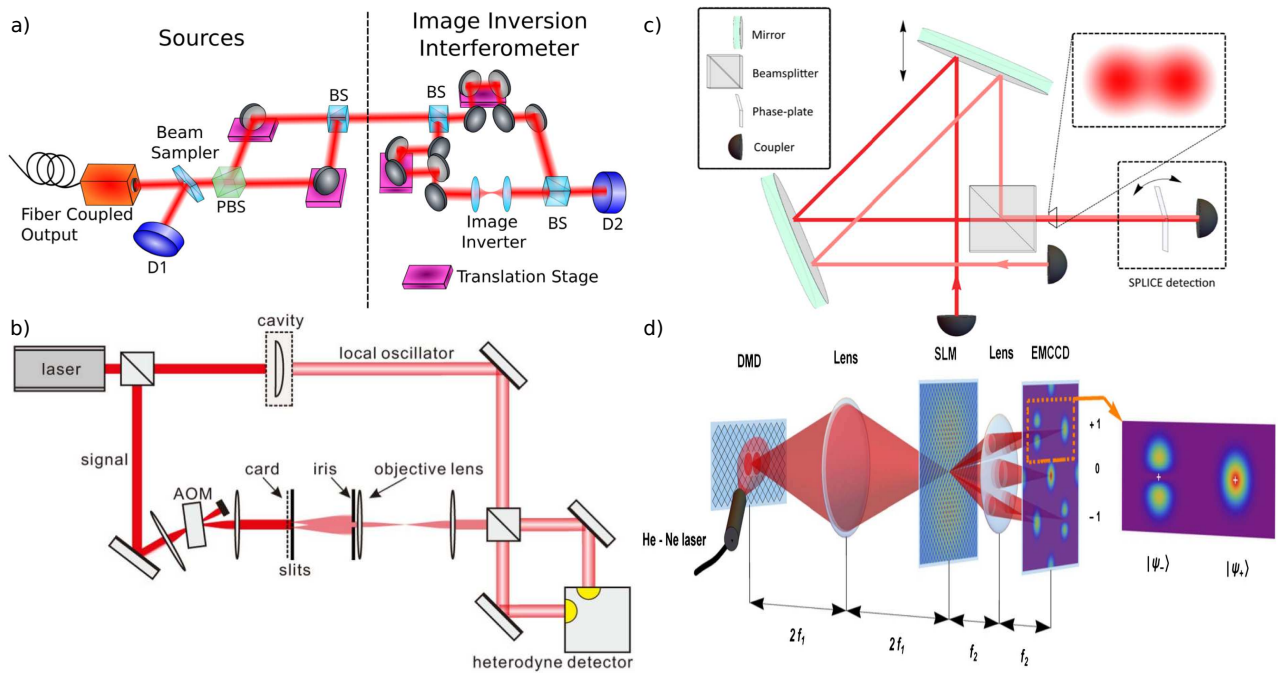}
\end{center}
\par
\vspace{-0.4cm}\caption{Proof-of-principle experiments demonstrating a quantum
detection scheme able to measure a distance of two incoherent point sources
better than the Rayleigh limit. \textbf{a)} Experiment of Tang et
al.\cite{Tang2016}. A HeNe laser with fiber coupled output is split at a
polarizing beamsplitter (PBS) into two beams of orthogonal polarization. They
are recombined at a beamsplitter (BS) with a slight lateral displacement to
simulate two incoherent light sources. The light sources are imaged using an
image inversion interferometer, a Mach-Zehnder interferometer with an image
inverter consisting of two lenses in one arm. One output of the interferometer
is detected by a photo detector (D2). \textbf{b)} Experiment of Yang et
al.\cite{Yang2016}. The signal beam is frequency shifted by an acousto-optical
modulator (AOM) and illuminates slits. A paper card is placed in front of the
slits to make the illumination incoherent. The signal beam is measured by
heterodyne detection using a local oscillator prepared in TEM01 mode by means
of an optical cavity. \textbf{c)} Experiment of Tham et al.\cite{Tham2017}.
Two partially overlapping beams as shown in the upper inset are generated by
coupling laser light out of a fiber and combining them on a beamsplitter. The
distance between the beams can be controlled by the position of the upper
mirror. The separation of the two beams is detected by projecting the beams
onto a mode orthogonal to TEM00, in their case a spatially antisymmetric field
mode. This is performed by passing the two beams through a phase plate which
is built in such a way that it introduces different phase shifts between
opposite halves of the beam and aligned such that coupling into a well-aligned
fiber coupler is minimized. The coupling into a single mode fiber corresponds
to a projection onto the TEM00 mode, thus together with the phase plate the
beams are projected onto a mode orthogonal to TEM00. \textbf{d)} Experiment of
Pa\'{u}r et al.\cite{Paur2016}. Using a high frequency switched digital micro
mirror chip (DMD) illuminated by a HeNe laser two closely spaced incoherent
beams are generated. The beam is projected onto different modes by an
amplitude spatial light modulator (SLM) generating a digital hologram. The
first-order diffraction spectrum is detected by an electron multiplying CCD
(EMCCD) camera. Figures adapted with permission from: a),
ref.~\onlinecite{Tang2016}, \copyright ~2016 OSA; b),
ref.~\onlinecite{Yang2016}, \copyright ~2016 OSA; c)
ref.~\onlinecite{Tham2017}, \copyright ~2017 APS; d)
ref.~\onlinecite{Paur2016}, \copyright ~2016 OSA.}%
\label{ExpRay}%
\end{figure*}

Shortly after the idea of Tsang et al.~\cite{Tsang15} was presented, it was
experimentally verified in several proof-of-principle experiments. The first
experiment by Tang et al.\cite{Tang2016} was based on super-localization by
image inversion interferometry\cite{Ran2016}. As shown in Fig.~\ref{ExpRay}%
(a), they used an image inversion interferometer to determine the separation
of two incoherent point sources, generated by two laser beams in orthogonal
polarizations stemming from the same HeNe laser. Using the light from the
simulated sources as the input, the interferometer was implemented as a
Mach-Zehnder interferometer with image inversion generated by a lens system in
one arm. The other arm was delayed appropriately so that the detector at the
output of the interferometer ideally showed no response for zero separation
due to destructive interference. With growing separation of the two sources
the destructive interference becomes more and more imperfect, yielding an
optical resolution beyond the Abbe-Rayleigh limit.

Yang et al.\cite{Yang2016} used heterodyne detection with a local oscillator
in TEM01 mode to detect the separation of the two slits in a double slit
configuration beyond the classical resolution limit. As also depicted in
Fig.~\ref{ExpRay}(b), they use a piece of paper to achieve incoherence and
diffuse transmission. Measuring at a frequency of some MHz to avoid noise at
lower frequencies, the beat between the local oscillator and the beam
illuminating the slits becomes zero if the separation is 0 as both spatial
parts of the TEM01 mode have a phase shift of $\pi$. Separating the two slits
yields a measurement beyond the Abbe limit. While the scheme requires the two
sources to be exactly aligned to the center of the TEM01 mode, using
higher-order TEM modes will provide general sub-Rayleigh imaging.

In another experiment, Tham et al.\cite{Tham2017} inserted a phase shift of
$\pi$ into the beam resembling the two incoherent sources (generated by
partially overlapping two beams in orthogonal polarizations) and projecting it
onto the guiding mode of a single mode fiber, as in Fig.~\ref{ExpRay}(c).
Finally, Pa\'{u}r et al.\cite{Paur2016} simulated Gaussian and slit apertures
by a digital micromirror chip illuminated by a laser. Projection onto
different modes was performed by a spatial light modulator which creates a
digital hologram measured by an electron-multiplying CCD. See
Fig.~\ref{ExpRay}(d). Let us conclude that super-resolving quantum imaging is
a hot topic and many other experiments could be
mentioned\cite{Treps,Gatto,Classen2016}.

\smallskip

\textbf{Discussion and outlook}

\noindent Quantum sensing is a rapidly evolving field with many potential
implications and technological applications. Despite the great advances that
have been achieved in recent years, a number of problems and experimental
challenges remain open. From the point of view of the basic theoretical models
of quantum metrology and quantum hypothesis testing, we may often compute the
ultimate performances allowed by quantum mechanics. However, we do not know in
general how to implement the optimal measurements achieving these performances
and/or what optimal states we need to prepare at the input of the unknown
quantum channel. Then, do we need to consider feedback and perform adaptive
protocols? For instance, this is an open question for both estimation and
discrimination of bosonic loss\cite{reviewMET}, which is at the basis of
quantum reading, quantum illumination and quantum-enhanced optical super-resolution.

From a more practical and experimental point of view, there are non-trivial
challenges as well. Despite a first proof-of-principle
demonstration\cite{Arno12} based on the unitary discrimination of
beamsplitters, we do not have yet a truly quantum reading experiment where a
single output of the cells is effectively accessed for the readout. A full
demonstration would involve an actual (one- or two-dimensional) array of
cells, where information is stored with classical codes and the quantum
readout is performed by simultaneously probing blocks of cells. This idea may
be further developed into a full experiment of bosonic quantum pattern
recognition where the use of entanglement across an array may boost the
resolution of unsupervised problems of data clustering.

Quantum illumination has had various experimental
demonstrations\cite{Qillexp1,Qillexp2,Qillexp3} but still remain the issue of
designing a practical receiver that\ would allow one to closely approximate
the theoretical limit forseen by the Helstrom bound. Challenges become more
serious when we consider the implementation of quantum illumination in the
microwave regime\cite{Qill2}. Here the development of highly-efficient
microwave-optical quantum converters could mitigate experimental issues
related with the generation of microwave entanglement and the detection of
microwave fields at the single-photon level. Furthermore these converters are
highly desired for other applications, in particular as interfaces for
connecting superconducting quantum chips and optical fibers in a potential
hybrid design of a future quantum Internet\cite{telereview}.

It is clear that other designs of a quantum radar are possible. For instance,
a fully microwave implementation of quantum illumination (without the use of
converters) may be achieved by using a superconducting Josephson parametric
amplifier to generate signal-idler microwave entanglement. Reflected signals
could then be phase-conjugated via another parametric amplifier, recombined
with the idlers, and finally measured, e.g., by using a transmon qubit as
single-photon detector. The idea of using Josephson mixers and photocounters
has been also studied by Las Heras et al.\cite{LasHeras} in the context of
using microwave quantum illumination to reveal phase-shift inducing cloaking,
also known as \textquotedblleft invisible cloak\textquotedblright.

Other experimental challenges need to be addressed in order to build an actual
quantum radar\cite{Qradar}. An important aspect is the preservation of the
idler modes while the signals are being propagated forward and back from the
target. The idlers should be kept in a low-loss delay line or stored in
quantum registers with sufficiently-long coherence times, until the final
joint detection. Then, unlike classical radars, whose performance improves as
the signal power is increased at constant bandwidth, the quantum counterpart
needs to increase bandwidth at constant signal brightness. The challenge is
therefore to generate microwave pulses with a time-bandwidth product of
$10^{6}$ modes or more. Furthemore, classical radars can interrogate many
potential target bins with a single pulse, while present models of quantum
radar may only query a single polarization, azimuth, elevation, range, Doppler
bin at a time. This is an area of development with very promising steps
forward\cite{Range}. Finally, more advanced technical issues related to
random-amplitude targets and radar clutters should also be addressed.

Regarding the experimental challenges for super-resolution\cite{Tang2016},
most of the current schemes, from spatial-mode demultiplexing to
super-localization by image inversion and heterodyne, rely on the assumption
that we need to know the location of the centroid of the sources in order to
get full quantum-optimal resolution. In general, this location is not exactly
known\cite{Tang2016,Yang2016,Tham2017,Paur2016}, so that achieving maximum
alignment before estimating the separation becomes an important step in order
to optimize the performance in a realistic implementation.

\smallskip

\textbf{Acknowledgments}

\noindent The authors would like to thank feedback from U. L. Andersen, L.
Banchi, Sh. Barzanjeh, J. Borregaard, S. L. Braunstein, V. Giovannetti, S.
Guha, C. Lupo, A. Lvovsky, M. Mikov\'{a}, M. Tsang, Z. Zhang. S.P. would like
to specifically thank J. H. Shapiro for discussions on the experimental
challenges related with a quantum radar, and R. Nair for the feedback on the
experimental challenges in optical super-resolution. S. P. thanks support from
the EPSRC via the `UK Quantum Communications Hub' (EP/M013472/1). T. G. would
like to acknowledge support from the Danish Research Council for Independent
Research (Sapere Aude 4184-00338B) as well as the Innovation Fund Denmark
(Qubiz) and the Danish National Research Foundation (Center for Macroscopic
Quantum States).


\bigskip

\begin{thebibliography}{999}                                                                                              %


\bibitem {Nielsen}Nielsen, M. A. \& Chuang, I. L. \textit{Quantum Computation
and Quantum Information} (Cambridge University Press, Cambridge, 2000).

\bibitem {Hayashi}Hayashi, M. \textit{Quantum Information Theory: Mathematical
Foundation} (Springer-Verlag, Berlin, 2017).

\bibitem {hybrid}U. L. Andersen, U. L., Neergaard-Nielsen, J. S., van Loock,
P. \& Furusawa, A. Hybrid discrete- and continuous-variable quantum
information. \textit{Nature Phys.} \textbf{11}, 713-719 (2015).

\bibitem {serale}Serafini, A. \textit{Quantum Continuous Variables: A Primer
of Theoretical Methods} (Taylor \& Francis, Oxford, 2017).

\bibitem {WeeRMP}Weedbrook, C. et al. Gaussian quantum information.
\textit{Rev. Mod. Phys.} \textbf{84}, 621 (2012).



\bibitem {SamPeter}Braunstein, S. L. \& Van Loock, P. Quantum information with
continuous variables. \textit{Rev. Mod. Phys.} \textbf{77}, 513 (2005).

\bibitem {Ger2}Adesso, G., Ragy, S. \& Lee, A. R. Continuous variable quantum
information: Gaussian states and beyond. \textit{Open Systems and Information
Dynamics} \textbf{21}, 1440001 (2014).

\bibitem {Degen}Degen, C. L., Reinhard, F. \& Cappellaro, P. Quantum sensing.
\textit{Rev. Mod. Phys.} \textbf{89}, 035002 (2017).

\bibitem {Sam1}Braunstein, S. L. \& Caves, C. M. Statistical distance and the
geometry of quantum states. \textit{Phys. Rev. Lett.} \textbf{72}, 3439 (1994).

\bibitem {Sam2}Braunstein, S. L., Caves, C. M. \& Milburn, G. J. Generalized
Uncertainty Relations: Theory, Examples, and Lorentz Invariance. \textit{Ann.
Phys.} \textbf{247}, 135-173 (1996).

\bibitem {SethSCIENCE}Giovannetti, V., Lloyd, S. \& Maccone, L.
Quantum-enhanced measurements: beating the standard quantum limit.
\textit{Science} \textbf{306}, 1330-1336 (2004).

\bibitem {Lore}Giovannetti, V., Lloyd, S. \& Maccone, L. Quantum
metrology.\ \textit{Phys. Rev. Lett.} \textbf{96}, 010401 (2006).

\bibitem {Paris}Paris, M. G. A. Quantum estimation for quantum technology.
\textit{Int. J. Quant. Inf.} \textbf{7}, 125-137 (2009).

\bibitem {Giova}Giovannetti, V., Lloyd S. \& Maccone, L. Advances in quantum
metrology. \textit{Nature Photon.} \textbf{5}, 222 (2011).

\bibitem {ReviewNEW}Braun, D. et al. Quantum enhanced measurements without
entanglement. Preprint at https://arxiv.org/abs/1701.05152 (2018).

\bibitem {Helstrom}Helstrom, C. W. \textit{Quantum Detection and Estimation
Theory} (Academic, New York, 1976).

\bibitem {QHT}Chefles, A. Quantum state discrimination. \textit{Contemp.
Phys.} \textbf{41}, 401 (2000).

\bibitem {QHT2}Barnett, S. M. \& Croke, S. Quantum state discrimination.
\textit{Adv. Opt. Photonics} \textbf{1}, 238-278 (2009).

\bibitem {UNA1}Bergou, J. A. , Herzog, U. \& Hillery, M.
\textit{Discrimination of Quantum States}, in Quantum state estimation,
Lecture Notes in Physics, \textbf{649}, 417-465 (Springer, Berlin, Heidelberg, 2004).

\bibitem {QCB1}Audenaert, K. M. R. et al. Discriminating States: The Quantum
Chernoff Bound. \textit{Phys. Rev. Lett.} \textbf{98}, 160501 (2007).

\bibitem {QCB2}Calsamiglia, J., Munoz-Tapia, R., Masanes, L., Acin, A. \&
Bagan, E., Quantum Chernoff bound as a measure of distinguishability between
density matrices: Application to qubit and Gaussian states. \textit{Phys. Rev.
A} \textbf{77}, 032311 (2008).

\bibitem {QCB3}Pirandola, S. \& Lloyd, S. Computable bounds for the
discrimination of Gaussian states. \textit{Phys. Rev. A} \textbf{78}, 012331 (2008).

\bibitem {QHB1}K. M. R. Audenaert, M. Nussbaum, A. Szkola, and F. Verstraete,
Commun. Math. Phys. \textbf{279}, 251 (2008).

\bibitem {Invernizzi}Invernizzi, C., Paris, M. G. A. \& Pirandola, S. Optimal
detection of losses by thermal probes. \textit{Phys. Rev. A} \textbf{84},
022334 (2011).

\bibitem {Gae1}Spedalieri, G. \& Braunstein, S. L. Asymmetric quantum
hypothesis testing with Gaussian states. \textit{Phys. Rev. A} \textbf{90},
052307 (2014).

\bibitem {reviewMET}Laurenza, R., Lupo, C., Spedalieri, G., Braunstein, S. L.
\& Pirandola, S. Channel Simulation in Quantum Metrology. Preprint at
https://arxiv.org/abs/1712.06603 (2017).

\bibitem {adaptivePRL}Pirandola, S. \& Lupo, C. Ultimate precision of adaptive
noise estimation. \textit{Phys. Rev. Lett. }\textbf{118}, 100502 (2017).

\bibitem {Prog1}Nielsen, M. A. \& Chuang, I. L. Programmable Quantum Gate
Arrays. \textit{Phys. Rev. Lett.} \textbf{79}, 321 (1997).

\bibitem {Prog2}Ji, Z., Wang, G., Duan, R., Feng, Y. \& Ying, M. Parameter
Estimation of Quantum Channels. \textit{IEEE Trans. Inform. Theory}
\textbf{54}, 5172--5185 (2008).

\bibitem {Prog3}Kolodynski, J. \& Demkowicz-Dobrza\'{n}ski, R. Efficient tools
for quantum metrology with uncorrelated noise. \textit{New J. Phys.}
\textbf{15}, 073043 (2013).

\bibitem {Prog4}Demkowicz-Dobrza\'{n}ski, R. \& Maccone, L. Using Entanglement
Against Noise in Quantum Metrology. \textit{Phys. Rev. Lett.} \textbf{113},
250801 (2014).

\bibitem {Harrow}Harrow, A. W., Hassidim, A., Leung, D. W. \& Watrous, J.
Adaptive versus nonadaptive strategies for quantum channel discrimination.
\textit{Phys. Rev. A} \textbf{81}, 032339 (2010).

\bibitem {PLOB}Pirandola, S., Laurenza, R., Ottaviani, C. \& Banchi, L.
Fundamental limits of repeaterless quantum communications. \textit{Nat.
Commun. }\textbf{8}, 15043 (2017). See also Preprint at
https://arxiv.org/abs/1510.08863 (2015).

\bibitem {TQC}Pirandola, S., Braunstein, S. L., Laurenza, R., Ottaviani, C.,
Cope, T. P. W., Spedalieri, G. \& Banchi, L. Theory of channel simulation and
bounds for private communication. Preprint at https://arxiv.org/abs/1711.09909 (2017).



\bibitem {Qread}Pirandola, S., Quantum Reading of a Classical Digital Memory.
\textit{Phys. Rev. Lett.} \textbf{106}, 090504 (2011).

\bibitem {Qread4}Lupo, C., Pirandola, S., Giovannetti, V. \& Mancini, S.
Quantum reading capacity under thermal and correlated noise. \textit{Phys.
Rev. A} \textbf{87}, 062310 (2013).

\bibitem {Qread3}Spedalieri, G., Lupo, C., Mancini, S., Braunstein, S. L. \&
Pirandola, S. Quantum reading under a local energy constraint. \textit{Phys.
Rev. A} \textbf{86}, 012315 (2012).

\bibitem {GaeENTROPY}Spedalieri, G. Cryptographic aspects of quantum reading.
\textit{Entropy} \textbf{17}, 2218-2227 (2015).

\bibitem {QreadCAP}Pirandola, S., Lupo, C., Giovannetti, V., Mancini, S. \&
Braunstein, S. L. Quantum Reading Capacity. \textit{New J. Phys.} \textbf{13},
113012 (2011).

\bibitem {LupoSUPER}Lupo, C. \& Pirandola, S. Super-additivity and
entanglement assistance in quantum reading. \textit{Quantum Inf. Comput.}
\textbf{17}, 0611 (2017).

\bibitem {SaikatJeff}Guha, S. \& Shapiro, J. H. Reading boundless error-free
bits using a single photon. \textit{Phys. Rev. A} \textbf{87}, 062306 (2013).

\bibitem {Saikat2}Guha, S., Dutton, Z., Nair, R., Shapiro, J. H. \& Yen, B.
Information Capacity of Quantum Reading in Laser Science, OSA Technical
Digest,Optical Society of America) paper LTuF2 (2011).

\bibitem {Nair11}Nair, R. Discriminating quantum-optical beam-splitter
channels with number-diagonal signal states: Applications to quantum reading
and target detection. \textit{Phys. Rev. A} \textbf{84}, 032312 (2011).

\bibitem {Yen11}Nair, R. \& Yen, B. J. Optimal Quantum States for Image
Sensing in Loss. \textit{Phis. Rev. Lett.} \textbf{107}, 193602 (2011)

\bibitem {Hirota11}Hirota, O., Error Free Quantum Reading by Quasi Bell State
of Entangled Coherent States. \textit{Quantum Measurements and Quantum
Metrology} \textbf{4}, 70-73 (2017).


\bibitem {Tej13}Prabhu Tej, J., Usha Devi, A. R. \& Rajagopal A. K. Quantum
reading of digital memory with non-Gaussian entangled light. \textit{Phys.
Rev. A} \textbf{87}, 052308 (2013).

\bibitem {Bisio11}Bisio, A., Dall'Arno, M. \& D'Ariano, G. M. Tradeoff between
energy and error in the discrimination of quantum-optical devices.
\textit{Phys. Rev. A} \textbf{84}, 012310 (2011).

\bibitem {Arno12}Dall'Arno, M., Bisio, A., D'Ariano, G. M., Mikov\'{a}, M.,
Je\v{z}ek, M. \& Du\v{s}ek, M. Experimental implementation of unambiguous
quantum reading. \textit{Phys. Rev. A} \textbf{85}, 012308 (2012).

\bibitem {ArnoIJQI}Dall'Arno, M., Bisio, A., D'Ariano, G.M. Ideal quantum
reading of optical memories. \textit{Int. J. Quant. Inf.} \textbf{10}, 1241010 (2012).

\bibitem {Qill0}Lloyd, S. Enhanced sensitivity of photodetection via quantum
illumination. \textit{Science} \textbf{321}, 1463 (2008).

\bibitem {Qill1}Tan, S.-H. et al. Quantum illumination with Gaussian States.
\textit{Phys. Rev. Lett.} \textbf{101}, 253601 (2008).

\bibitem {JeffNJP}Shapiro, J. H. \& Lloyd, S. Quantum illumination versus
coherent-state target detection. \textit{New J. Phys.} \textbf{11}, 063045 (2009).

\bibitem {JeffRECEIVER}Zhuang, Q., Zhang, Z. \& Shapiro, J. H. Optimum
Mixed-State Discrimination for Noisy Entanglement-Enhanced Sensing.
\textit{Phys. Rev. Lett.} \textbf{118}, 040801 (2017).

\bibitem {ROC}Zhuang, Z., Zhang, Z. \& Shapiro, J. H. Entanglement-enhanced
Neyman--Pearson target detection using quantum illumination. \textit{J. Opt.
Soc. Am. B} \textbf{34}, 1567 (2017).

\bibitem {Qill2}Barzanjeh, Sh. et al. Microwave quantum illumination.
\textit{Phys. Rev. Lett.} \textbf{114}, 080503 (2015).

\bibitem {SaikatREC}Guha, S. \& Erkmen, B. I. Gaussian-state
quantum-illumination receivers for target detection. \textit{Phys. Rev. A}
\textbf{80}, 052310 (2009).

\bibitem {QillAMP}Xiong, B., Li, X., Wang, X-.Y. \& Zhou, L. Improve microwave
quantum illumination via optical parametric amplifier. \textit{Ann. Phys.}
\textbf{385}, 757-768 (2017).

\bibitem {Qillmetro}Sanz, M., Las Heras, U., Garc\'{\i}a-Ripoll, J. J.,
Solano, E. \& Di Candia, R. Quantum estimation methods for quantum
illumination. \textit{Phys. Rev. Lett.} \textbf{118}, 070803 (2017).

\bibitem {Qill3}Weedbrook, C., Pirandola, S., Thompson, J., Vedral, V. \& Gu,
M. How discord underlies the noise resilience of quantum illumination.
\textit{New J. Phys.} \textbf{18}, 043027 (2016).

\bibitem {Ragy}Ragy, S. et al., Quantifying the source of enhancement in
experimental continuous variable quantum illumination, \textit{J. Opt. Soc.
Am. B} \textbf{31}, 2045--2050 (2014).

\bibitem {QillASY1}Wilde, M. M., Tomamichel, M., Lloyd, S., \& Berta, M.
Gaussian Hypothesis Testing and Quantum Illumination. \textit{Phys. Rev.
Lett.} \textbf{119}, 120501 (2017).

\bibitem {QillASY2}De Palma, G. \& Borregaard, J. The ultimate precision of
quantum illumination. arXiv:1802.02158 (2018).

\bibitem {Qillexp1}Lopaeva et al. Experimental realization of quantum
illumination. \textit{Phys. Rev. Lett.} \textbf{110}, 153603 (2013).

\bibitem {Meda}Meda, A., Losero, E., Samantaray, N., Scafirimuto, F.,
Pradyumna, S., Avella, A., Ruo-Berchera, I. \& Genovese, M. Photon-number
correlation for quantum enhanced imaging and sensing. \textit{J. Opt.}
\textbf{19}, 094002 (2017).

\bibitem {Qillexp2}Zhang, Z., Tengner, M., Zhong, T., Wong, F. N. C. \&
Shapiro, J. H. Entanglement's benefit survives an entanglement-breaking
channel. \textit{Phys. Rev. Lett.} \textbf{111}, 010501 (2013).

\bibitem {Qillexp3}Zhang, Z., Mouradian, S., Wong, F. N. C. \& Shapiro, J. H.
Entanglement-enhanced sensing in a lossy and noisy environment. \textit{Phys.
Rev. Lett.} \textbf{114}, 110506 (2015).

\bibitem {LasHeras}Las Heras, U., Di Candia, R., Fedorov, K. G., Deppe, F.,
Sanz, M. \& Solano, E. Quantum illumination reveals phase-shift inducing
cloaking. \textit{Sci. Rep.} \textbf{7}, 9333 (2017).

\bibitem {Qradar}M. Lanzagorta, \textit{Quantum Radar (Synthesis Lectures on
Quantum Computing)}, Vol. 3, No. 1, pp. 1-139 (Morgan \& Claypool, 2011).

\bibitem {Tsang15}Tsang, M., Nair, R. \& Lu, X.-M. Quantum theory of
superresolution for two incoherent optical point sources. \textit{Phys. Rev.
X} \textbf{6}, 031033 (2016).

\bibitem {Lupo16}Lupo, C. \& Pirandola, S. Ultimate precision bound of quantum
and subwavelength imaging. \textit{Phys. Rev. Lett.} \textbf{117}, 190802 (2016).

\bibitem {Tsang2}Nair, R. \& Tsang, M. Far-Field Superresolution of thermal
electromagnetic sources at the quantum limit. \textit{Phys. Rev. Lett.}
\textbf{117}, 190801 (2016).

\bibitem {Kerviche}Kerviche, R., Guha, S. \& Ashok, A. Fundamental limit of
resolving two point sources limited by an arbitrary point spread function.
Preprint at https://arxiv.org/abs/1701.04913 (2017).

\bibitem {Rehacek}Rehacek, J., Pa\'{u}r, M., Stoklasa, B., Hradil, Z. \&
S\'{a}nchez-Soto, L. L. Optimal measurements for resolution beyond the
Rayleigh limit. \textit{Opt. Letters} \textbf{42}, 231-234 (2017).

\bibitem {Yang17}Yang, F., Nair, R., Tsang, M., Simon, C. \& Lvovsky, A. I.
Fisher information for far-field linear optical superresolution via homodyne
or heterodyne detection in a higher-order local oscillator mode. \textit{Phys.
Rev. A} \textbf{96}, 063829 (2017).

\bibitem {Lu2018}Lu, X.-M., Krovi, H., Nair, R., Guha, S., Shapiro, J. H.
Quantum-optimal detection of one-versus-two incoherent optical sources with
arbitrary separation, arXiv:1802.02300 (2018).

\bibitem {Tang2016}Tang, Z. S., Durak, K. \& Ling, A. Fault-tolerant and
finite-error localization for point emitters within the diffraction limit.
\textit{Opt. Express} \textbf{24}, 22004 (2016).

\bibitem {Ran2016}Nair, R. \& Tsang, M. Interferometric superlocalization of
two incoherent optical point sources. \textit{Opt. Express} \textbf{24},
3684-3701 (2016).

\bibitem {Yang2016}Yang, F., Taschilina, A., Moiseev, E. S., Simon, C. \&
Lvovsky, A. I. Far-field linear optical superresolution via heterodyne
detection in a higher-order local oscillator mode. \textit{Optica} \textbf{3},
1148 (2016).

\bibitem {Tham2017}Tham, W. K., Ferretti, H. \& Steinberg, A. M. Beating
Rayleigh's Curse by Imaging Using Phase Information. \textit{Phys. Rev. Lett.}
\textbf{118}, 070801 (2017).

\bibitem {Paur2016}Pa\'{u}r, M., Stoklasa, B., Hradil, Z., Sanchez-Soto, L. L.
\& \v{R}eh\'{a}\v{c}ek, J. Achieving the ultimate optical resolution.
\textit{Optica} \textbf{3}, 1144-1147 (2016).


\bibitem {Gatto}Gatto Monticone, D. et al. Beating the Abbe Diffraction Limit
in Confocal Microscopy via Nonclassical Photon Statistics. \textit{Phys. Rev.
Lett.} \textbf{113}, 143602 (2014).

\bibitem {Treps}Treps, N., Andersen, U. L., Buchler, B., Lam, P. K.,
Ma\^{\i}tre, A., Bachor, H.-A. \& Fabre, C. Surpassing the Standard Quantum
Limit for Optical Imaging Using Nonclassical Multimode Light. \textit{Phys.
Rev. Lett.} \textbf{88}, 203601 (2014).

\bibitem {Classen2016}Classen, A. et al. Superresolving imaging of arbitrary
one-dimensional arrays of thermal light sources using multiphoton
interference. \textit{Phys. Rev. Lett.} \textbf{117}, 253601 (2016).



\bibitem {Tsang2009}Tsang, M. Quantum imaging beyond the diffraction limit by
optical centroid measurements. \textit{Phys. Rev. Lett.} \textbf{102}, 253601 (2009).

\bibitem {Rozema2014}Rozema, L. A. et al. Scalable spatial superresolution
using entangled photons. \textit{Phys. Rev. Lett.} \textbf{112}, 223602 (2014).



\bibitem {qcomb}Chiribella, G., D'Ariano, G. M. \& Perinotti, P. Quantum
circuit architecture. \textit{Phys. Rev. Lett.} \textbf{101}, 060401 (2008).

\bibitem {teleBennett}Bennett, C. H. et al. Teleporting an unknown quantum
state via dual classical and Einstein-Podolsky-Rosen channels. \textit{Phys.
Rev. Lett.} \textbf{70}, 1895 (1993).

\bibitem {Samtele}Braunstein, S. L. \& Kimble, H. J. Teleportation of
continuous quantum variables. \textit{Phys. Rev. Lett.} \textbf{80}, 869-872 (1998).

\bibitem {telereview}Pirandola, S., Eisert, J., Weedbrook, C., Furusawa, A \&
Braunstein, S. L. Advances in quantum teleportation. \textit{Nat. Photonics}
\textbf{9}, 641-652 (2015).

\bibitem {Monras}Monras, A. \& Paris, M. G. A. Optimal quantum estimation of
loss in bosonic channels. \textit{Phys. Rev. Lett.} \textbf{98}, 160401 (2007).

\bibitem {Banchi}Banchi, L., Braunstein, S. L. \& Pirandola S. Quantum
fidelity for arbitrary Gaussian states. \textit{Phys. Rev. Lett.}
\textbf{115}, 260501 (2015).

\bibitem {AdessoOPT}Adesso, G., Dell'Anno, F., Siena, S. D., Illuminati, F. \&
Souza, L. A. M. Optimal estimation of losses at the ultimate quantum limit
with non-Gaussian states. \textit{Phys. Rev. A} \textbf{79}, 040305(R) (2009).



\bibitem {Fuchs}Fuchs, C. A. \& van de Graaf, J. Cryptographic
distinguishability measures for quantum-mechanical states. \textit{IEEE Trans.
Inf. Theory} \textbf{45}, 1216 (1999).

\bibitem {Prepres}Sudarshan, E. C. G. Equivalence of Semiclassical and Quantum
Mechanical Descriptions of Statistical Light Beams. \textit{Phys. Rev. Lett.}
\textbf{10}, 277 (1963).

\bibitem {Prepres2}Glauber, R. J. Coherent and Incoherent States of the
Radiation Field. \textit{Phys. Rev.} \textbf{131}, 2766 (1963).

\bibitem {Discord}Modi, K., Brodutch, A., Cable, H., Paterek, T. \& Vedral, V.
The classical-quantum boundary for correlations: Discord and related measures.
\textit{Rev. Mod. Phys.} \textbf{84}, 1655 (2012).

\bibitem {EOM1}Barzanjeh, Sh., Vitali, D., Tombesi, P. \& Milburn, G. J.
Entangling optical and microwave cavity modes by means of a nanomechanical
resonator. \textit{Phys. Rev. A} \textbf{84}, 042342 (2011).

\bibitem {EOM2}Barzanjeh, Sh., Abdi, M., Milburn, G. J., Tombesi, P. \&
Vitali. Reversible Optical-to-Microwave Quantum Interface. D. \textit{Phys.
Rev. Lett.} \textbf{109}, 130503 (2012).

\bibitem {EOM3}Andrews, R. W. et al. Bidirectional and efficient conversion
between microwave and optical light. \textit{Nat. Phys.} \textbf{10}, 321 (2014).



\bibitem {Range}Zhuang, Q., Zhang, Z. \& Shapiro, J. H. Entanglement-enhanced
lidars for simultaneous range and velocity measurements. \textit{Phys. Rev. A}
\textbf{96}, 040304(R) (2017).
\end{thebibliography}
\end{document}